\documentclass[%
mathleft,%
final,%
]{an}
\usepackage{graphicx}
\usepackage{times}
\begin{document}

\Pagespan{1}{}
\Yearpublication{2013}%
\Yearsubmission{2012}%
\Month{1}%
\Volume{334}%
\Issue{1}%
\DOI{This.is/not.aDOI}%

\title{Multiple views of magnetism in cool stars}

\author{%
J.~Morin\inst{1}\fnmsep\thanks{Corresponding author:\email{jmorin@gwdg.de}}\and%
M.~Jardine\inst{2}\and %
A.~Reiners\inst{1}\and %
D.~Shulyak\inst{1}\and %
B.~Beeck\inst{1}\and %
G.~Hallinan\inst{3}\and %
L.~Hebb\inst{4}\and %
G.~Hussain\inst{5}\and %
S.~V.~Jeffers\inst{1}\and %
O.~Kochukhov\inst{6}\and %
A.~Vidotto\inst{2}\and %
L.~Walkowicz\inst{7} %
}
\authorrunning{T.H.E. Editor \& G.H. Ostwriter}
\institute{
Institut f\"ur Astrophysik, 
Georg-August-Universit\"at G\"ottingen, Friedrich-Hund Platz,
37077 G\"ottingen, Germany 
\and
School of Physics and Astronomy,
Univ.\ of St~Andrews, St~Andrews,
Scotland KY16 9SS, UK
\and
California Institute of Technology,
1200 E. California Blvd., Pasadena,
CA 91125, USA
\and
Department of Physics and Astronomy,
Vanderbilt University, Nashville,
TN 37235, USA
\and
ESO,
Karl-Schwarzschild-Str. 2,
D-85748 Garching, Germany
\and
Department of Physics and Astronomy,
Uppsala University Box 516,
751 20 Uppsala, Sweden
\and
Department of Astrophysical Sciences,
Princeton University, Princeton,
NJ 08544, USA
}

\received{XXXX}
\accepted{XXXX}
\publonline{XXXX}


\abstract{%
  Magnetic fields are regarded as a crucial element for our understanding of
  stellar physics. They can be studied with a variety of methods which provide
  complementary -- and sometimes contradictory -- information about the
  structure, strength and dynamics of the magnetic field and its role in the
  evolution of stars. Stellar magnetic fields can be investigated either with
  direct methods based on the Zeeman effect or through the observation of
  activity phenomena resulting from the interaction of the field with the
  stellar atmosphere.
  In this Cool Stars \textsc{XVII} Splinter Session we discussed
  the results obtained by the many ongoing studies of stellar activity and
  direct studies of surface magnetic fields, as well as the state-of-the-art 
  techniques on which they are based. We show the strengths and
  limitations of the various approaches currently used and to point out
  their evolution as well as the interest of coupling various magnetism and
  activity proxies.
}

\maketitle

\section{Introduction}
Since the first detection on the Sun through the Zeeman effect
in 1908 by G.~E. Hale, magnetic fields have become a key
element of stellar physics. They are now thought to play an important role
throughout the formation and evolution of stars and their planetary systems.
In addition, they are now detected across a large part of the H-R diagram (e.g.
Donati \& Landstreet 2009; Reiners 2012),
either directly or through the activity phenomena they power which can be
observed across a large part of the electromagnetic spectrum. The
characterization of these magnetic fields has revealed that stars show a very
wide variety of magnetic properties in terms of strength, geometry or
variability. Drawing a clear picture of stellar magnetism and dynamo processes
is however not easy since different types of measurements do not always agree --
at least apparently -- and have led to various interpretations (e.g. Reiners et
al. 2009; Morin et al. 2011 in the case of M dwarfs).

In this session, we discussed a number of approaches
used to detect, measure and model stellar magnetic fields as well as
the related activity phenomena. A particular emphasis was put on the
comparison of different approaches or proxies, the way they can be combined, and
on the ongoing developments in instrumentation, analysis and modelling.

\section{Surface magnetic field and starspots}
\subsection{Starspots from spectroscopy and spectropolarimetry}
\label{sec:zdi}
Direct measurements of stellar magnetic fields at the photospheric level rely on
the properties of the Zeeman effect. Two complementary techniques are
successfully employed. On the one hand, Zeeman-induced line broadening measured
in unpolarized light directly provides a measurement of the average
surface field modulus -- weighted e.g. by limb-darkening and brightness
inhomogeneities.
This approach can be applied regardless of the field complexity, but it only
provides very little information about the field geometry.
Such measurements are limited to low $v\sin{i}$ values (up to $\sim20~{\rm
km/s}$).
On the other hand, the measurement of polarisation in spectral lines is
sensitive to the vector properties of the surface magnetic field. However this
technique can only probe the large-scale structure of stellar magnetic fields
-- due to the cancellation of polarised signal corresponding to neighbouring
regions of opposite polarities -- and is blind to bipolar groups for instance.

Doppler imaging (DI) is a powerful tool for investigating long-lived stellar
surface features on rapidly-rotating stars (Vogt \& Penrod 1983). It
relies on the correspondence between the spatial position of a surface
inhomogeneity on a rotating star and the wavelength of the corresponding
distortion in the Doppler-broadened line profile. It allows to retrieve the
locations, shapes and sizes of starspots. M.~Semel (1989) first proposed to
extend DI to polarised spectra in order to investigate magnetic fields
and termed this new technique ``Zeeman-Doppler Imaging'' (ZDI). In particular,
ZDI can disentangle between field orientations, and allowed to
demonstrate the existence of magnetic fields quite different from the Sun's,
for instance in young solar-type stars with shallow convective envelopes
(Jeffers \& Donati 2008).

\subsection{Seeing Spots: Observations of Stellar Magnetic Activity with Kepler}
Although Kepler's primary mission is the detection of extrasolar planets, its
incredible precision and high cadence provide an exquisite opportunity to study
the interplay between stellar rotation -- both global and differential -- and
magnetic activity.  These observations allow us to place our Sun in the context
of other stars, and improve not only our understanding of the stars themselves,
but the range of environments in which planets exist, a crucial element to
model their surfaces and atmospheres.

The Kepler mission has now obtained over three years of nearly
contiguous observations for Sun-like stars. These lightcurves allow to
determine rotation periods, measure differential rotation and detect flares on a
wide variety of stars, both similar to and much cooler than the Sun. With such a
wealth of new information it is now possible to make new advances on the
rotation--activityrelation across the H-R diagram (e.g. Basri et al. 2011); or
to derive stellar rotation axis inclinations (when
combined with $v\sin{i}$ measurements, Buchave et al. \textit{in prep.}).
The combination of rotation period measurements with asteroseismic diagnostics
derived from Kepler ligh-curves also bears the potential of extending
gyrochronology calibrations to slowly rotating stars. Ongoing efforts to
characterize stellar variability and activity phenomena in the Kepler target
stars include modelling of spot distributions, detections of differential
rotation and flare statistics (e.g., Walkowicz et al. 2011; Walkowicz, Basri \& 
Valenti, \textit{submitted}).

\subsection{Understanding the effect of magnetic fields on the fundamental
properties of low mass stars through ZDI and eclipsing binary
modelling of YY~Gem}
A multi-faceted analysis of the magnetic field properties and global
fundamental parameters of the well known M dwarf eclipsing binary YY~Gem
(e.g. Torres \& Ribas 2002) is ongoing. Understanding the
magnetic field properties of YY~Gem is important because it is one of the few
systems that provide a benchmark for stellar evolution models of early M
dwarfs, and may allow us to understand how activity affects stellar structure
(see e.g. Hebb et al. 2012).

Through a coordinated observing campaign, coincident high precision multi-band
light curves and time series spectropolarimetry of the system were obtained over
10 consecutive nights in January 2012. Preliminary magnetic field maps and
surface brightness maps derived from the ZDI analysis of both
components of the system were derived. The resulting surface brightness maps are
incorporated as constraints on the starspot parameters when completing a new
eclipsing binary modelling. In this way, precise and accurate masses and radii
could be derived for both stars. The updated value of the discrepancy between
observed and theoretical mass--radius relation is about 5~\%, lower than
previous estimates. Yet more eclipsing binary systems must be analysed to
to solve this long-standing problem of mass--radius relation in active low-mass
stars.

Finally, a preliminary analysis of the frequency and flux of the optical flares
that were observed during the photometric campaign was performed. By combining
these multiple analyses, a first global picture of the large-scale magnetic
field properties of two early M dwarf stars can be drawn. This study ultimately
aims to assess the effect of the magnetic field on the fundamental
parameters of cool stars. 

\subsection{Can we map starspots and their magnetic fields using infra-red
lines and molecular bands?}
Currently employed methods of spectropolarimetric analysis of cool active stars
-- such as Least-Squares Deconvolution combined with ZDI
(cf.~Sec.~\ref{sec:zdi}) -- are systematically missing  strong magnetic
fields inside cool starspots since they do not take into account temperature
inhomogeneities. It is widely believed that an extension of polarimetric studies
to molecular bands and magnetically sensitive infra-red lines may allow
detection and mapping of starspot magnetic fields. Magnetometry in the near
infra-red is indeed attractive due to the wavelength dependence of the Zeeman
effect and the reduced spot-to-photosphere intensity contrast. Atomic lines in
the near infra-red however tend to be shallower than in the visible.
Zeeman-sensitive molecular lines -- mainly TiO ($7055~$\AA) and FeH
($9900~$\AA) -- on their side can prove valuable to probe magnetic
fields inside cool spots, although the amplitude polarization signal in these
lines is lower than in Fe~\textsc{i} lines (see e.g. Shulyak et al. 2010).

These possibilities are tested with comprehensive ZDI simulations based on the
methodology presented by Kochukhov et al. (2009). The calculations demonstrate
that neither molecular features nor infra-red atomic lines taken separately
provide sufficient information to simultaneously recover the contrast and
geometry of starspots. Both diagnostics must be combined with optical lines in
order to achieve substantial improvement of the magnetic and temperature
reconstruction.

\subsection{MHD simulations of surface convection in cool main-sequence stars}
\label{sec:bbeeck}
Cool main-sequence stars have thick convective envelopes or are fully
convective. Magnetic fields have been detected in many of such stars. In the
Sun, the observed surface magnetic field is highly structured owing to its
interaction with the convective flows. The local structure of the magnetic field
in other cool stars is, however, unknown. In the absence of
spatially resolved observations, the effect of the local structure of the
magnetic field on detectable signals like magnetically sensitive spectral lines
can be evaluated by numerical simulations of the magneto-convective processes.

Comprehensive 3D radiative magnetohydrodynamic simulations of the surface layers
of main-sequence stars of spectral types F3 to M2 were carried out (see e.g.
V\"ogler et al. 2012). The simulations were analysed in terms of
sizes and properties of the convection cells (granules) and magnetic flux
concentrations. Synthetic spectral line profiles were generated and the impact
of the different magnetic structures of these stars on their spectra was
investigated.

These simulations show a qualitative difference in magneto-convection between
solar-like stars and M dwarfs (Beeck et al. 2011). The
synthetic spectra yield rough lower limits
for the detectability of magnetic fields in cool stars depending on rotation
velocity and global field geometry. It is shown that the net effect of magnetic
field -- broadening, distortion and apparent radial velocity shift -- on a
Zeeman-sensitive spectral line is largely degenerate with the star's $v\sin{i}$
and radial velocity for average field moduli up to 500~G (50~mT). Fitting
several spectral lines simultaneously will likely help overcoming this issue.
Finally, these results can be used to test magnetic field measurement methods
like ZDI, because they provide known field configurations and their
time-dependent line spectra. 

\section{Coronae and magnetospheres}

\subsection{Understanding stellar coronae: Insights and challenges}
Surface magnetic field maps of cool stars have been used to produce detailed
models of the coronae and extended environments of cool stars that can
successfully reproduce a range of observational characteristics. These coronal
field models are powerful tools that have the potential to explain how a cool
star system evolves from the pre-main sequence -- when the star is actively
accreting from its primordial disk -- to the main sequence -- when magnetically
driven stellar winds control the angular momentum evolution of stars.

Stellar coronae can be studied through their X-ray emission
both in quiescent and flaring states. Rotational modulation of X-ray emission
can be detected in active binaries, and the X-ray-emitting coronae of active
G and K dwarfs are shown to be very compact. Current stellar coronal models are
generally based on extrapolations of the surface magnetic maps reconstructed
with ZDI -- thus defining the footpoints of X-ray emitting coronal loops -- and
assume an isothermal corona in hydrostatic equilibrium. X-ray emission can then
be computed assuming an optically thin corona. Such models have been successful
at reproducing X-ray rotational modulation (e.g. Hussain et al. 2007) as well as
constraining the location of accretion hotspots
in T Tauri stars (e.g. Gregory et al. 2010).

One of the challenges faced by such models is the limited spatial resolution of
ZDI surface magnetic maps, and their impact on the derived quantities 
(e.g. Arzoumanian et al. 2011;
Jardine et al. \textit{in prep.}). New methods based on MHD dynamical
modelling allow to take into consideration self-consistently the interaction of
the outflowing wind with the magnetic field and vice versa (Vidotto et al.
2011). Upcoming observational capabilities such as ALMA will
also prove useful in probing circumstellar environments and enabling future
advances.

\subsection{Radio Signatures of Magnetospheric Phenomena on Active Stars, Brown
Dwarfs, and Extrasolar Planets}
The radio wavelength domain offers powerful tools to characterize stellar
and planetary magnetic fields. The actual derivation of magnetic field
strength depends on the emission process involved.
For the Sun, quiescent
emission is dominated by thermal emission from the chromosphere and the corona,
whereas non-thermal emission is powered by particle acceleration during
impulsive events like flares. Planets on their side produce a very bright
low-frequency (kHz--MHz) emission powered by quasi-stable particle acceleration.
VLA observations revealed that many active stars also
possess a nonthermal corona where large populations of
electrons are continuously accelerated to high energies, a situation with no
solar counterpart.

The advent of a new generation of radio observatories, together with the 
upgrades to existing facilities, offers the potential for significant
breakthroughs in the study of magnetospheric phenomena on active stars, brown
dwarfs and exoplanets as well as new avenues for synergies with observations in
other wavelength regimes. Interferometers, such as the Jansky VLA, now offer the
capability of time-resolved wideband spectroscopy enabling the detection and
frequency characterization of sweeping radio bursts, a capability
previously only available in observations of the Sun.

Future multi-wavelength
campaigns will allow the correlation of such bursts with the higher energy
counterparts from optical to X-rays. Very Long Baseline interferometry (VLBI)
enables the resolved imaging of coronal structures on nearby stars during both
quiescence and flare events (e.g. Peterson et al. 2010),
which in turn can be directly mapped to the large scale magnetic fields revealed
by tomographic techniques such as ZDI. Beyond the main
sequence, the detection of auroral, pulsed radio emission provides the sole
means for direct determination of magnetic field strengths for cool brown
dwarfs (e.g. Hallinan et al. 2007), exemplified by the
recent radio detection of a T6.5 dwarf (Route \& Wolszczan 2012). Similar
emissions may soon be detectable from extrasolar planets with the advent of the
new generation of low frequency arrays, such as
the LWA and LOFAR. 

\subsection{Investigating the effects of magnetic cycles on the Tau Boo system}
To quantify the effect stellar cycles have on orbiting planets, one has to
understand the properties of the stellar wind, which depend on the particular
geometry of the coronal magnetic field at each epoch during the stellar cycle.
$\tau$~Boo is an intriguing planet-host star that is believed to undergo
magnetic cycles similar to the Sun, but with a duration that is about one order
of magnitude smaller than that of the solar cycle. With the use of
observationally derived surface magnetic field maps, the magnetic stellar wind
of $\tau$~Boo can be studied by means of three-dimensional MHD numerical
simulations (Vidotto et al. 2012).

In total, four surface magnetic maps are used, they encompass one full magnetic
cycle i.e. two polarity reversals (e.g. Fares et al. 2009).
Wind mass loss-rate is found to vary little during the observed epochs of the
cycle (less than 3 per cent), with a relatively more important variation in
angular momentum loss-rates (a factor of 2). The models show
that the emission measure from the quiescent closed-field line corona remains
approximately constant through the cycle, in agreement with recent X-ray
observations of $\tau$~Boo (Poppenhaeger et al. 2012). This rather
cycle-independent X-ray emission from $\tau$~Boo could point to the presence of
overlapping activity cycles (e.g., overlap in the butterfly diagram, McIvor et
al. 2006).

The rather rapid changes in the magnetic properties of the stellar
corona and the relatively high mass-loss rates of $\tau~Boo$ ($\sim 2.7\times
10^{-12}$~M$_\odot\,$yr$^{-1}$, i.e., two orders of magnitude larger than that
of the solar wind) imply that the environment that surrounds $\tau$~Boo~b (and
other putative planets in this system) is significantly different from the
environment surrounding the planets in the solar system. The radio emission
from the hot Jupiter that orbits at 0.0462~au from $\tau$~Boo is estimated.
It is shown that, for a planet with a magnetic field similar to
Jupiter ($\sim14$~G at the pole), the radio flux is estimated to be about
1.4~mJy, occurring at a frequency of 34 MHz, potentially detectable from the
ground.

\section{Summary and discussion}
\label{sec:summary}
The interpretation and comparison of magnetic field measurements and modelling
stemming from different approaches is a debated topic. There is, however, a
general agreement that a better knowledge of the spatial structure of
stellar magnetic fields is essential in order to reconcile these different
views, and that what we know about the Sun likely cannot be safely extrapolated
to stars with very different properties (in particular very active stars). For
instance, the predominance of bright (plage-like) \textit{vs} dark
(spot-like) regions among magnetic surface features, the very existence of
regions corresponding to ``quiet'' photosphere on very active stars, are of
great relevance to magnetic field measurements and start to be addressed
through numerical simulations (sec.~\ref{sec:bbeeck}). This importance of the
spatial structure of the field is even relevant to the vocabulary used.
It can be argued that ``magnetic field'' should be reserved for an actual
magnetic field vector at some point of the stellar surface, whereas
observationally-derived quantities should be termed ``magnetic flux''. Both
quantities have the same unit of Gauss (CGS) or Tesla (SI), although some
confusion can arise with a quantity also called ``magnetic flux'' and measured
in Maxwell (i.e. G.cm$^2$, CGS) or Weber (i.e. T.m$^2$, SI), see Reiners (2011)
and Reiners \& Mohanty (2012). 

The talks given in the session suggest that in the next few years progress
will arise from the use of new or upgraded observational facilities,
coupled with improved modelling methods and from multi-wavelength or
multi-technique studies. Ultimately, such advances will aim at a better
characterization of stellar magnetic fields and planetary environments, as
well as a better understanding of magnetic field generation in cool stars.

\acknowledgements
  We thank the organizers of the 17th Cool Stars conference for giving us the
  opportunity to hold this session. JM acknowledges the support
  of the Alexander von Humboldt foundation.

%

\end{document}